\documentclass[final,3p,times,twocolumn]{elsarticle}

\usepackage{amssymb}
\usepackage{pdflscape}
\usepackage{gensymb}
\usepackage[modulo]{lineno}
\usepackage{caption}
\usepackage{subcaption}
\usepackage{graphicx}
\usepackage{float}
\usepackage{booktabs}
\usepackage{stackengine}
\usepackage{amssymb}
\usepackage[applemac]{inputenc}
\usepackage[T1]{fontenc}
\usepackage{xcolor}

\journal{Nuclear Instruments and Methods in Physics Research A}

\begin{document}

\begin{frontmatter}



\title{Recent Upgrades of the Gas Handling System for the Cryogenic Stopping Cell of the FRS Ion Catcher}


\cortext[mycorrespondingauthor]{Corresponding authors}
\author[inst1,inst2,inst3]{A.~Mollaebrahimi\corref{mycorrespondingauthor}}\ead{Ali.Mollaebrahimi@exp2.physik.uni-giessen.de}
\author[inst2]{D.~Amanbayev}
\author[inst3]{S.~Ayet San Andr\'{e}s}
\author[inst2,inst3]{S.~Beck}
\author[inst2]{J.~Bergmann}
\author[inst2,inst3]{T.~Dickel}
\author[inst2,inst3]{H.~Geissel}
\author[inst2,inst3]{C.~Hornung}
\author[inst1]{N.~Kalantar-Nayestanaki}
\author[inst2]{G.~Kripko-Koncz}
\author[inst2]{I.~Miskun}
\author[inst4,inst5]{D.~ Nichita}
\author[inst2,inst3]{W.~R.\ Pla\ss}
\author[inst3,inst6]{I.~Pohjalainen}
\author[inst2,inst3,inst7]{C.~Scheidenberger}
\author[inst8]{G.~Stanic}
\author[inst4,inst5]{A.~State}
\author[inst3]{J.~Zhao}

\affiliation[inst1]{organization={Nuclear Energy Group, ESRIG, University of Groningen},
            addressline={Zernikelaan 25}, 
            city={Groningen},
            postcode={9747 AA}, 
            state={Groningen},
            country={The Netherlands}}

\affiliation[inst2]{organization={II.~Physikalisches Institut, Justus-Liebig-Universit{\"a}t Gie{\ss}en},
            addressline={Heinrich-Buff-Ring}, 
            city={Gie{\ss}en},
            postcode={35392}, 
           state={Hessen},
           country={Germany}}

\affiliation[inst3]{organization={GSI Helmholtzzentrum f{\"u}r Schwerionenforschung GmbH},
            addressline={Planckstra{\ss}e 1}, 
            city={Darmstadt},
            postcode={64291}, 
           state={Hessen},
           country={Germany}}

\affiliation[inst4]{organization={Extreme Light Infrastructure-Nuclear Physics (ELI-NP)},
            addressline={Reactorului 30}, 
            city={Bucharest},
            postcode={077125}, 
           state={Bucharest},
           country={Romania}}

\affiliation[inst5]{organization={Doctoral School in Engineering and Applications of Lasers and Accelerators, University Polytechnica of Bucharest},
            addressline={Splaiul Independentei 313}, 
            city={Bucharest},
            postcode={060811}, 
           state={Bucharest},
           country={Romania}}

 \affiliation[inst6]{organization={Department of Physics, University of Jyvaskyla},
            addressline={P.O. Box 35 (YFL)}, 
            city={Jyvaskyla},
            postcode={FI-40014}, 
           state={Jyvaskyla},
           country={Finland}}          

\affiliation[inst7]{organization={Helmholtz Research Academy Hesse for FAIR (HFHF), GSI Helmholtz Center for Heavy Ion Research},
            addressline={Heinrich-Buff-Ring 16}, 
            city={Gie{\ss}en},
            postcode={35392}, 
           state={Hessen},
           country={Germany}}

\affiliation[inst8]{organization={German Cancer Research Center (DKFZ)},
            addressline={Im Neuenheimer Feld 280}, 
            city={Heidelberg},
            postcode={69120}, 
           state={Heidelberg},
           country={Germany}}           
           
\begin{abstract}
In this paper, the major upgrades and technical improvements of the buffer gas handling system for the cryogenic stopping cell of the FRS Ion Catcher at GSI/FAIR (in Darmstadt, Germany) are described. The upgrades include implementation of new gas lines and gas purifiers to achieve a higher buffer gas cleanliness for a more efficient extraction of reactive ions as well as suppression of the molecular background ionized in the stopping cell. Furthermore, additional techniques have been implemented for improved monitoring and quantification of the purity of the helium buffer gas.
\end{abstract}




\end{frontmatter}


\section{Introduction}
\label{intro}

Efficient production, separation and identification of radioactive nuclei is one of the most challenging and important steps for research with exotic nuclei \cite{Blumenfeld_2013}. At in-flight facilities, projectile fragmentation and fission processes are used to produce exotic nuclei at high kinetic energies and therefore, several steps are required to deliver purified and cooled low-energy beams for high-precision measurements (mass spectrometry, decay and laser spectroscopy). First, the produced exotic nuclei are separated by the \mbox{$B \rho - \Delta E - B \rho$} method \cite{FRS-facility}. Subsequently, they are slowed down by energy loss in degraders \cite{FRS-facility}. Finally, they are fully stopped and thermalized in a gas-filled stopping cell (typically helium) via buffer gas collisions. The low-energy ions are then extracted from the stopping cell by means of DC and RF fields and gas flow to the downstream measurement setups \cite{WADA2013450}. Stopping cells have been continuously developed and improved \cite{NEUMAYR2006489,LUND2020378, cryo, Ranjan_2011,super-frscell,Chouhan_2014,WADA2013450,RIKEN-cell}, since their first application about two decades ago \cite{RIKEN-cell,WEISSMAN2005245}. A main factor limiting their performance is the cleanliness of the buffer gas, as it strongly influences the ion survival and the extraction efficiency of ions from the stopping cell \cite{KUDRYAVTSEV2001412,cryo}. In addition it defines the background contamination delivered to the downstream setups \cite{ELISEEV2007479}.

\begin{figure*} [h]  
\centering
\includegraphics[scale=0.52]{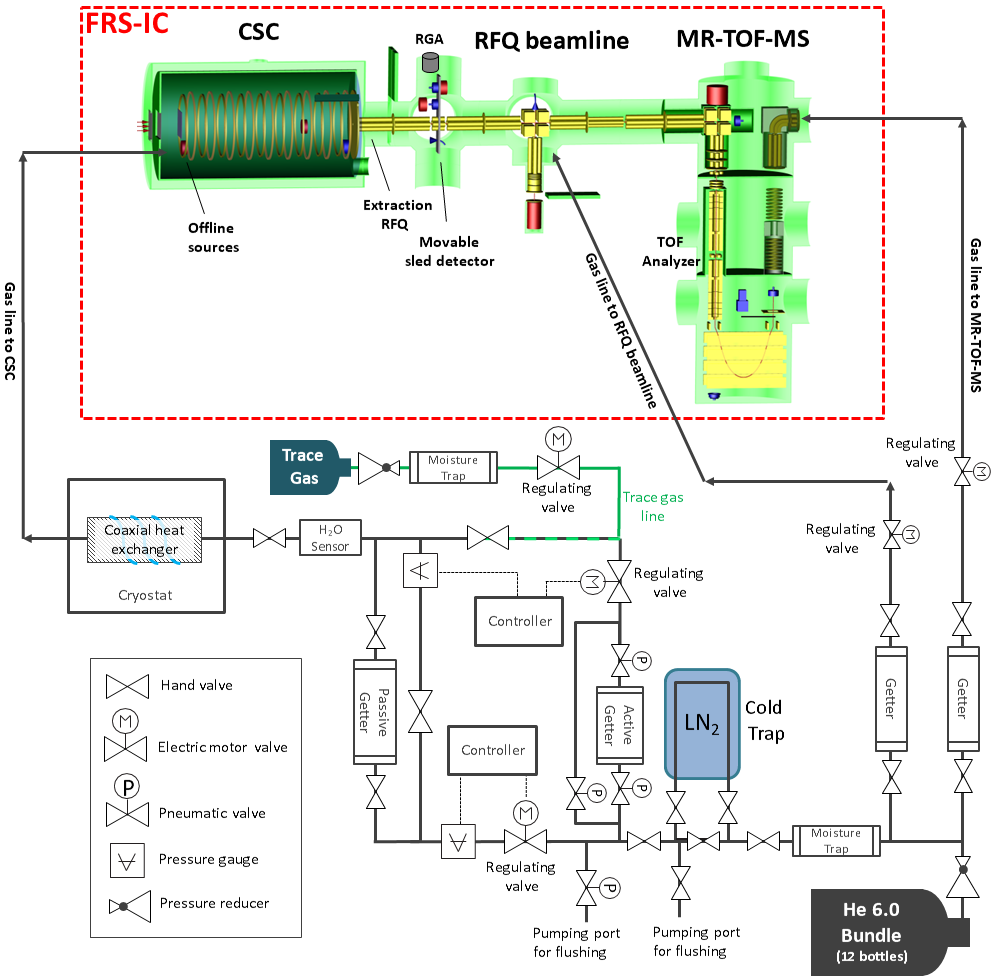}
\caption{The upgraded gas-handling system of the FRS-IC setup is shown. The FRS-IC setup is on the top side. The helium gas source (bundle) is shown in the bottom right corner. The gas can pass through the different gas lines for filling the CSC and is also used for the RFQ beamline and the MR-TOF-MS trap systems. Several commercial gas purifiers along with an in-house made cold trap are used for purification of the helium gas. An additional trace-gas line allows to add trace amounts of gases to the main helium gas going to the CSC.}
\label{gas-handeling}
\end{figure*}

Contamination in the buffer gas can result in the formation of molecules by the chemical reactions of the ions of interest with the impurities in the gas or in charge-exchange reactions and even neutralization of the ions of interest. The formation of molecular ions can result in a significant change of their ion mobility, changing their extraction efficiency by the RF carpet \cite{Ivan-IJMS}. Moreover, the ions of interest can be distributed over several charge or molecular states, thus making their measurement more difficult and less efficient. The impurities in the gas cell are ionized by charge-exchange reactions with buffer gas ions, which themselves were created by the interaction of the fast beam with the buffer gas. These impurities result in additional unwanted space-charge at the RF carpet nozzle as well as in the downstream experiments, e.g., the high-precision mass-spectrometer, reducing the ion-transport efficiency. Extracted impurities can overlap with the ions of interest with similar mass-to-charge ($A/q$) values, thus limiting the sensitivity of the measurement. Therefore, great efforts are made to ensure a high cleanliness of the buffer gas. One of the techniques to achieve this is the use of a gas purifier in the gas-inlet lines of the stopping cell \cite{pascal,cell,KALEJA2020280,KUDRYAVTSEV20084368}.

At the GSI Helmholtz Center for Heavy Ion Research, in Germany, research with exotic nuclei and their applications is done. The FRagment Separator (FRS) is the key-device for this research. It is a high-resolution in-flight spectrometer and separator for nuclides produced in projectile fragmentation or fission reactions \cite{FRS-facility}. The FRS Ion Catcher (FRS-IC) \cite{FRS-wolfgang2013} is located at the final focal plane of the symmetric branch of the FRS. Here, the high-energy fragments are slowed down and thermalized for high-precision measurements \cite{Wplass2019-science}. The main parts of the FRS-IC are the Cryogenic Stopping Cell (CSC) \cite{cryo,Purushothaman_2013,pascal} for stopping and thermalization of the ions in the gas, the RFQ beamline \cite{Christine-thesis, pascal, Emma,IDI-Florian} for cooling, identification, and transport of the low-energy ions to the Multiple-Reflection Time-of-Flight Mass Spectrometer (MR-TOF-MS) \cite{Timo-MRTOF,PLA20084560}, which is used for the high-precision mass spectrometry and isobaric/isomeric separation \cite{Timo-PLB}. The FRS-IC setup also serves as a test facility for similar experiments at the Super-FRS of the upcoming FAIR facility \cite{super-frss, super-frscell}. The following sections describe the status and improvements of the gas handling system of the FRS-IC setup.

\section{The buffer-gas cleanliness of the CSC and its recent improvements}

The CSC of the FRS-IC has a factor two higher areal gas density (product of density and length) than comparable systems worldwide \cite{Wplass2019-science,LUND2020378}. This makes the FRS-IC setup much more sensitive to gas impurities due to the higher number of collisions between the ions and the buffer gas atoms. The gas in the CSC is expected to have nitrogen gas as the main contamination, as it is a specified contaminant in the helium gas used, helium 6.0 (99.9999\% purity), and the gas cleaning methods applied so far do not remove it. The charge-states of the extracted elements fit with the ionization potential corresponding to N$_2$. \cite{Ivan-Phd}. About 70\% of the elements are extracted as singly-charged ions and about 30\% as doubly-charged ions due to the different ionization potentials. The extraction of ions with high first-ionization potentials (Ne, F and Ar) will not be possible as long as there is a significant N$_2$ contamination in the buffer gas. 

Besides the cleanliness of the gas supplied to the CSC, the residual gas in the CSC before filling it with helium is equally important. To ensure low base pressure, the CSC is not only operated at cryogenic temperature, but before cooling down also baked out with a set temperature of 125$^{\degree}$C ($\pm 5^{\degree}$C) for at least 48 hours. This results in 80$^{\degree}$C for the coldest recorded area (RF carpet). Considering a gas cell operating at 100 mbar, 80 K and an average extraction time of 25 ms \cite{FRS-wolfgang2013} the ions undergo $\approx 10^8$ collisions with the buffer gas atoms before they are extracted. Thus, a gas cleanliness of 1 part per billion (ppb) is needed to ensure maximum ion survival and extraction efficiency from the stopping cell. 


Figure \ref{gas-handeling} depicts the upgraded gas-distribution system of the FRS Ion Catcher. The slow-control system for monitoring, controlling and logging of the status of the cryogenic and pressure information of the CSC has also been upgraded and is presented in Ref. \cite{STATE2022166772}. Highest gas quality is achieved by employing gas purifiers, the liquid nitrogen cold trap, the collision cleaning of the gas line (purging) \cite{collision-cleaning}, and only using UHV compatible components in the gas handling system.  The recent upgrades include new gas lines for a more flexible gas handling system, the cold trap filled with charcoal for the suppression of heavy noble gases, a heated active getter (SAES Monotorr purifier), the helium bundles with larger capacity for long experiments and a completely new trace-gas line for the charge-state manipulations of the ions in the CSC. The cold trap is the first stage in the gas cleaning process, as it can be easily regenerated and thus limits the amount of contaminants going to the later stages of the gas cleaning process.

\section{Diagnostics methods} \label{diognostic}

The MR-TOF-MS is the primary method to monitor the ionized impurities extracted from the CSC. The MR-TOF-MS has a high mass resolving power and can distinguish between different isobaric contaminants. But it measures the ions only after they have been transported through the beamline and stored in several gas-filled RF traps, thus after possible dissociation or charge-exchange reactions in the beamline. Also measuring ions with a mass-to-charge ratio $m/q<40$ u/e is challenging with the MR-TOF-MS as the RF frequencies in the MR-TOF's RFQs are at the moment optimized for the mass-to-charge range of 40 to 240 u/e. 
Therefore, additional techniques have been developed and implemented for identification, quantification, and monitoring of the impurities in the stopping gas. Moreover, these allow the CSC to have its own diagnostics, independent of the MR-TOF-MS. The techniques allow the monitoring of the neutral contaminations present in the gas, as well as the ionized contamination extracted from the CSC. 

\subsection{Extraction RFQ operated as mass filter}
The extraction RFQ can be used to measure the ionized impurities extracted from the CSC, when it is operated as an RF mass filter \cite{Misku2015, IDI-Florian}. In this mode, the transmitted ions are detected by a channeltron detector or a Si detector mounted on a movable detector sled in the RFQ beamline. The channeltron detector is operated in the conversion-dynode mode to increase the detection efficiency. The data acquisition and control of the mass filter is described in Ref. \cite{Emma}. 

\subsection{Residual gas analyzer (RGA)}
The Residual-Gas Analyzer (RGA) is a commercial quadrupole gas analyzer (Hiden Analytical HMT100) installed on the RFQ beamline and operating at room temperature, see Figure \ref{gas-handeling}. It is technically not possible to monitor the gas condition directly inside CSC since the maximum operating pressure of a typical RGA amounts to about $10^{-4}$ mbar. However, the RGA can be used to monitor the gas coming from CSC through the nozzle to the differential pumped RFQ section downstream (Figure \ref{gas-handeling}).


\section{Cold-trap gas purifier}
The cold trap is a gas purifier developed in-house and used in addition to the commercial gas purifiers (SAES Microtorr and Monotorr) for further improvement of the gas cleanliness. The main goal of the cold trap is the suppression of the noble gases with low vapor pressures at liquid-nitrogen temperature (i.e., Xe and Kr). The cold trap is a dewar filled with liquid nitrogen (LN$_2$) and the gas line is passed through the LN$_2$. The cold section of the gas line is filled with activated charcoal and operated at about 2 bar helium gas pressure. The choice to use the activated carbon is made due to the easier regeneration at lower temperatures (100$^{\degree}$ C) compared to other materials like zeolite.

Suppression of impurities in the cold trap is done by surface adsorption on the activated charcoal, which is a very powerful adsorbent due to its huge surface area ($\approx10^3$ m$^2$/g), used widely in industrial applications. The method has also been used with other stopping cells \cite{NEUMAYR2006489,Ilka}. Cooling of the adsorbents materials to cryogenic temperature can also massively increase the adsorption capacity. The cryogenic operation of the CSC itself can also suppress contaminants, but the cryogenic surface of the CSC is quickly saturated. The 85~g of charcoal in the cold trap has a surface area of around $8.5\times10^4$ m$^2$, which is about four orders of magnitude larger than the surface of the CSC. Moreover, the activated charcoal can be easily and quickly (a few hours) regenerated if it is saturated. It is typically saturated in a few weeks depending on the cleanliness of the gas and being less effective for adsorption of impurities. Thus, the cold trap is usually regenerated every two weeks of permanent operation. The regeneration is done by baking out the charcoal while pumping the gas line.


\subsection{Xe and Kr suppression with the cold trap}
Although the Xe and Kr contamination levels are not specified in the impurity compositions of the helium 6.0 gas bottles, Xe and Kr can be present in the gas due to the history of the gas bottles. The gas bottles are long-lived components, which are refilled and reused many times for different gases in their lifetime. Therefore, a trace amount of noble gases can be present in the gas bottle filled with helium. When the FRS-IC is operated in similar condition, the observed Xe contamination in two arbitrary helium gas bottles changes from a factor of 0.6 to 4 compared to fragments measured from a $^{252}$Cf spontaneous fission source located inside the CSC \cite{252Cf,252Cf2}. The Xe atoms are ionized by the charge-exchange reaction with helium ions and then they are extracted from the CSC and identified by the MR-TOF-MS.



Suppressing these noble gases (e.g., Kr and Xe) is important for mass measurements with the FRS Ion Catcher, as they generate large backgrounds at many mass numbers. We measured noble-gas ions forming adducts with the impurities in the He gas (e.g.~KrN$_2^+$ or XeO$_2^+$) in the clean cryogenic environment of the CSC. Furthermore, compared to the usual light impurities, such as H$_2$O and N$_2$, the noble gas impurities are more difficult to remove using mass spectrometric methods, since they have mass-to-charge values similar to those of medium-heavy exotic nuclei.

\begin{figure}[H]
\begin{subfigure}{1\linewidth}
\centering
\phantomcaption
{\includegraphics[width=1\linewidth]{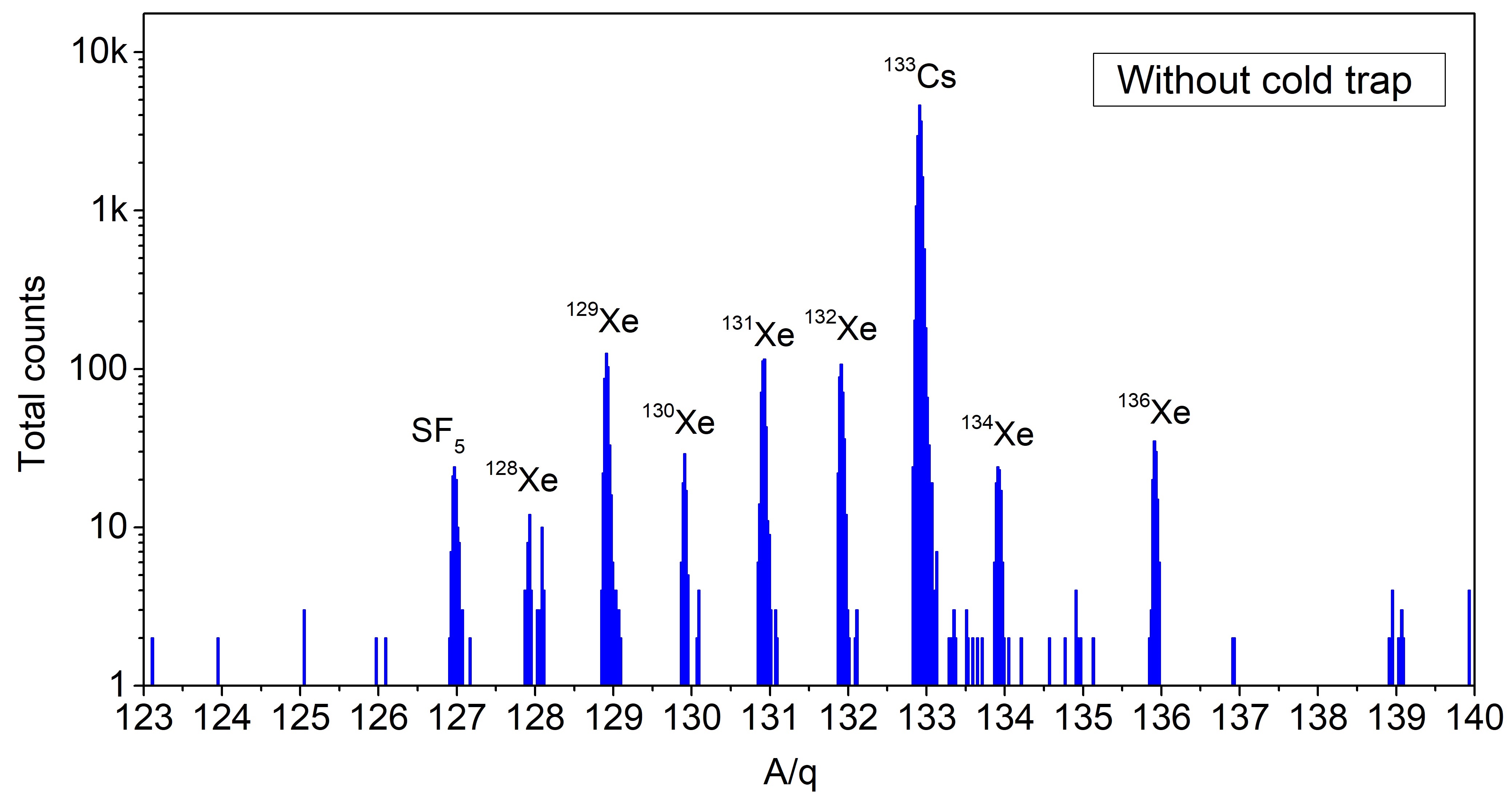}}
\end{subfigure}%
\vspace{-0.6cm}
\begin{subfigure}{1\linewidth}
\centering
\phantomcaption
 {\includegraphics[width=1\linewidth]{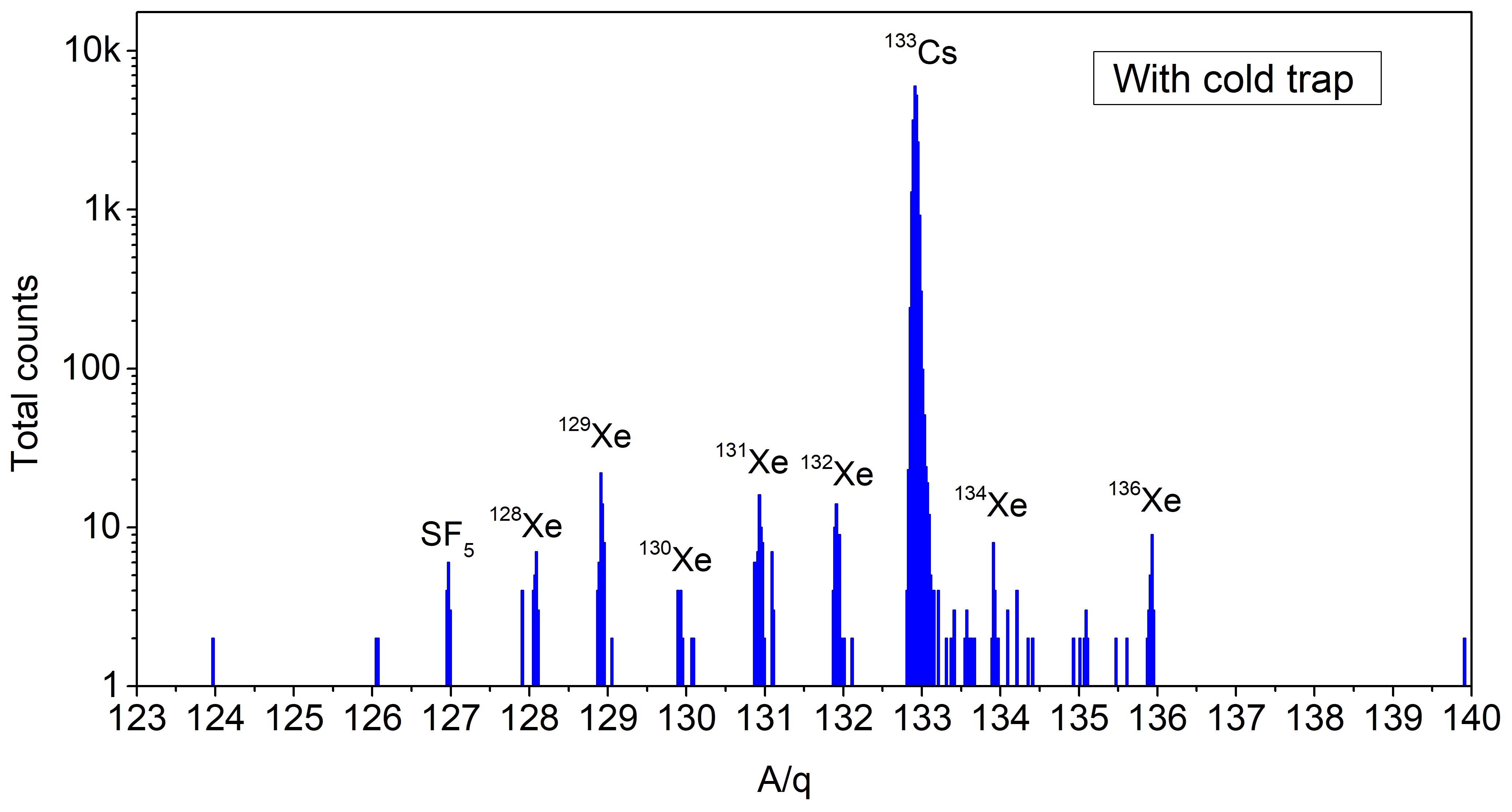}}
\end{subfigure}
\captionsetup{width=1.0\linewidth, justification=justified}
\caption[]{Mass spectra of ions extracted from the CSC filled without (top panel) and with (bottom panel) the cold trap in the gas line. The measurements were performed using the MR-TOF-MS.}
\label{Xe}
\end{figure}

To investigate the suppression of the noble gases, CSC was filled with 50 mbar helium with and without the cold trap in the gas line. The gas lines and the CSC were pumped before filling the CSC in each measurement to reach a similar base pressure. The extracted count rate of Xe$^+$ was then monitored with the MR-TOF-MS in both conditions. The gas was ionized by a $^{252}$Cf fission source and a $^{228}$Th $\alpha$-source installed inside CSC with a nominal activity of 37 kBq and 1.4 kBq, respectively. 
 Table \ref{Xe-table} (column a) shows the ion count rate for the most natural abundant xenon isotope ($^{132}$Xe$^+$) with and without the cold trap. A suppression factor of 7.3 is observed for using the cold trap. The same behavior was also found for the other Xe isotopes with smaller natural abundances. Figure \ref{Xe} shows the logarithmic mass spectrum for the same measurement with and without the cold trap for all Xe isotopes. The dominant $^{133}$Cs peak is the calibration peak from internal ion source of the MR-TOF-MS. SF$_5$ and $m/q=139$ u/e are other impurities coming from the gas, which are also decreased when using the cold trap.

\begin{figure}[H] 
\centering
\includegraphics[width=1\linewidth]{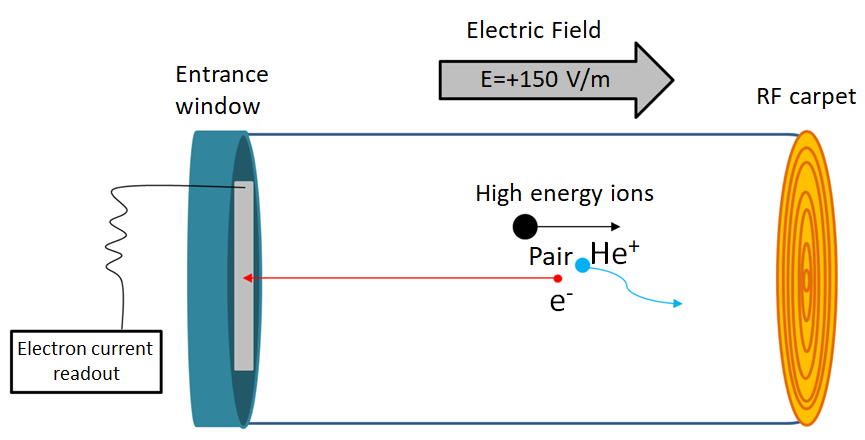}
\captionsetup{width=0.9\linewidth, justification=justified}
\caption[]{Electron-current measurement on the entrance window of the CSC. The current is created by the deposited energy from stopping of the high-energy ions inside the CSC.}
\label{entrance}
\end{figure}

The rates of Xe$^+$ can be much higher with heavy fragments from the FRS impinging on the CSC. On average, about $10^7$ He$^+$e$^-$ pairs are generated for each heavy nuclide \cite{REITER2016240} which do not recombine due to the fast removal of electrons in presence of the applied strong electric field. To estimate the Xe concentration in the gas and the expected Xe$^+$ rate during such an online measurement one needs to measure the ionization during the here-presented offline measurements \cite{Dragos-thesis}. This can be done by measuring the electron current on the entrance window of the DC cage in the CSC. A weak electric field is applied to collect the electrons produced during the stopping of the high-energy ions; see Figure \ref{entrance}.

The electron current was measured for different pressures in offline tests while the gas is ionized by the two radioactive sources installed inside the CSC ($^{228}$Th and $^{252}$Cf). The higher the pressure, the higher the stopping power of the gas, resulting in a higher current on the entrance flange, until the saturation point (more than 300 mbar) is reached when all ions from the sources are stopped in the gas. 
A current of 33 pA was measured at a gas pressure of 50 mbar, which corresponds to production of about $2 \times 10^8$ He$^+$e$^-$ pairs per second. As each He$^+$ undergoes about $1 \times 10^8$ buffer gas collisions before extraction or neutralization one gets a total number of collisions per second of $2\times10^{16}$. By detecting the Xe$^+$ ions in the MR-TOF-MS and considering the extraction, transport and the detection efficiencies of 4\% one can calculate the number of Xe$^+$ produced inside the CSC. 
This is done by assuming that each collision of a He$^+$ with a Xe atom results in a charge-exchange reaction and thus the production of a Xe$^+$. A charge-exchange reaction rate of 7$\times10^{-12}$ cm$^3$/s is reported for He$^+$ in Xe gas and it is expected to be a few orders of magnitude larger for dimer and trimer helium ions (existing in a cryogenic environment) based on the experimental data of the helium dimer (He$_2^+$) in room temperature noble gases like Ar or Kr \cite{Anicich2003AnIO}. Therefore, we estimate the Xe concentration in the helium gas (Xe/He ratio) by the ratio of produced Xe$^+$ to the total number of collisions (shown in column b of Table \ref{Xe-table}). The detection limit is below $10^{-17}$. This enormous sensitivity may have unique applications in ultra-trace gas analysis \cite{MURTZ2006963}. Assuming the typical rate of $ 10^4 $ s$^{-1}$ fragments impinging on the CSC, the expected number of Xe$^+$ reaching the MR-TOF-MS (table \ref{Xe-table} column c) was calculated. The MR-TOF-MS is typically operated with a repetition frequency of 50~Hz. To achieve highest mass accuracy, ion-ion interaction should be avoided when ions are flying in the MR-TOF-MS analyzer; thus a maximum of 1 ion per cycle and per mass number should be measured. One can see that the suppression of Xe with the cold trap reduced the Xe level below that threshold; enabling mass measurement of exotic nuclei also for the mass numbers of Xe.

Kr atoms are much more difficult to remove than Xe atoms due to their lower vapor pressure at liquid nitrogen temperature. Furthermore, the measured Kr level was much lower as compared to Xe. Therefore, a discharge source \cite{Ivan-Phd} was used inside the CSC during the Kr measurements to facilitate ionization and thus get a significant count rate. Thus, no evaluation could be done for the Kr concentration and expected rates during the online measurements due to the unknown amount of ionization produced by using the discharge source (see table \ref{Xe-table}).

\begin{table*}[h]  
\small
\captionsetup{width=0.95\textwidth, justification=justified}
\caption{Column a: the Xe$^+$ and Kr$^+$ count rates coming from the CSC filled with and without the cold trap in the gas line; column b: the Xe/He contamination level in the gas cell (see text for details); column c: the expected Xe$^+$ count rate identified in the MR-TOF-MS during the online measurements assuming $10^4$ s$^{-1}$ fragments stopping in the gas on average; the last column shows the improvement factor, on average, by using the cold trap in the gas line. 
}
\scriptsize
\centering
\begin{tabular}{ c|ccc|ccc|c  }
\toprule
 &    & \textbf{Without cold trap} &&  & \textbf{With cold trap} && \\
\midrule
&a&b&c&a&b&c&\\
\midrule
&&  & Beam && & Beam &\\
Isotope    & Identified  & [Xe/He]  &  Ionization & Identified & [Xe/He] &  Ionization  & Improvement \\
 &  (counts/s) &   &  (counts/s) & (counts/s) & &  (counts/s) & factor\\
\midrule
$^{132}$Xe &  0.58& $7.0 \times 10^{-16}$ &280& 0.08 & $9.5 \times 10^{-17}$ &38& 7.3\\
\midrule
$^{84}$Kr &  0.83& - &-& 0.40 & - &-& 2.1\\
\midrule
\end{tabular}
\label{Xe-table}
\end{table*}



\begin{figure}[h]
\begin{subfigure}{1\linewidth}
\centering
 \phantomcaption
{\includegraphics[width=1\linewidth]{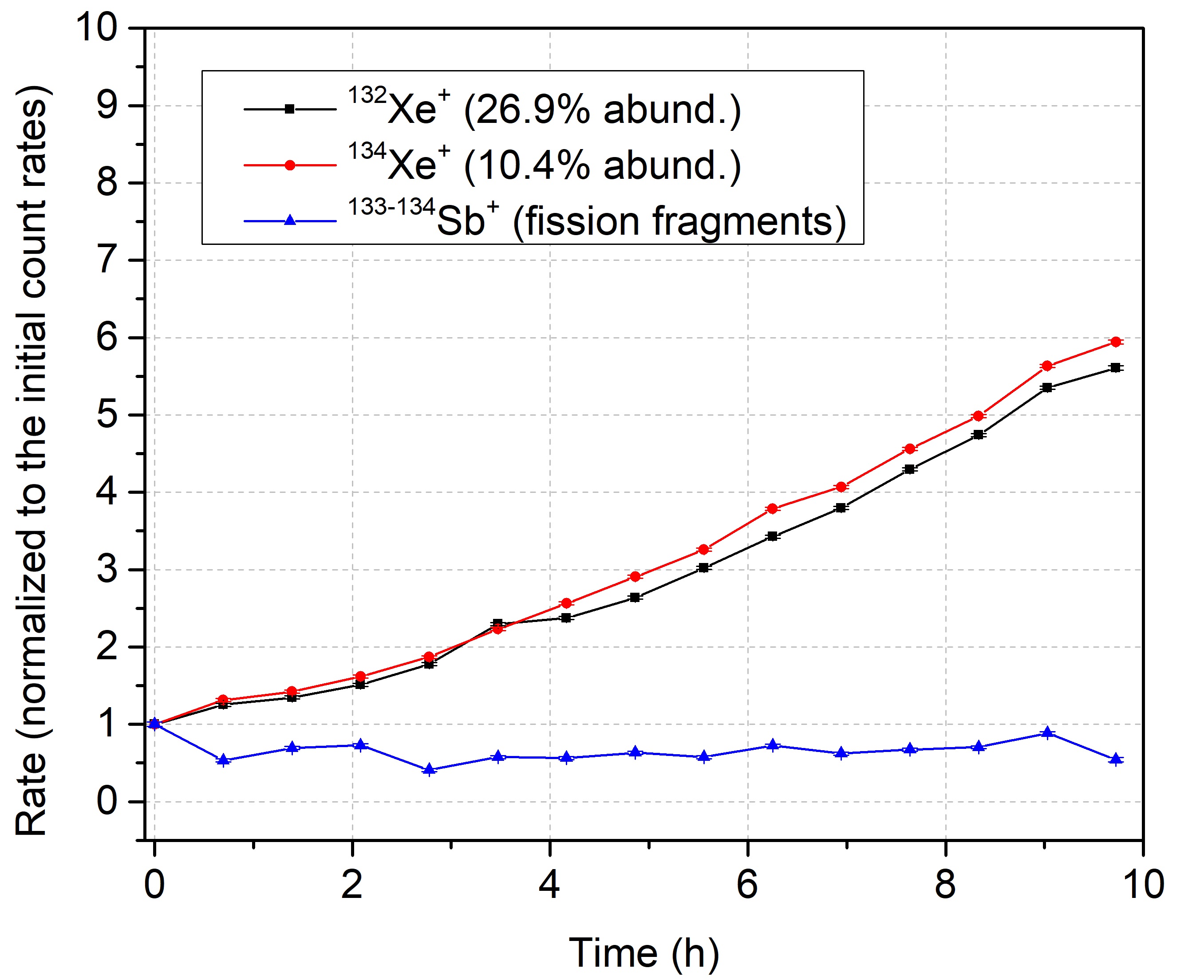}}
\end{subfigure}
\captionsetup{width=1.0\linewidth, justification=justified}
\caption[]{The ion rates for $^{132}$Xe$^+$ and $^{134}$Xe$^+$ coming from the impurities of the gas without using the cold trap in the gas line and the Sb fission fragments ($A=133-134$) from the $^{252}$Cf source. The rates are normalized to the initial count rate to show the change in time.}
\label{Xe-ff-nocoldtrap}
\end{figure}

The suppression of Xe atoms by the cold trap was also studied over longer time periods (i.e., days). The Xe count rates together with fission fragments at $A/q=132-134$ (from the $^{252}$Cf source installed inside the CSC) were measured over about 10 hours with the MR-TOF-MS. The CSC was operated at 165 mbar without the cold trap in the gas line. A high pressure of 165 mbar was needed for efficient stopping and extraction of the high-energy fission fragments from the $^{252}$Cf source. 
It was observed that the Xe rate was increasing by about a factor six, whereas the rates of the fission fragments were stable (Figure \ref{Xe-ff-nocoldtrap}). The increase of Xe rate is expected to be due to saturation of the cold surface inside the CSC for adsorption of Xe atoms.

\begin{figure}[h] 
\centering
\hspace{0 cm}\includegraphics[width=1\linewidth]{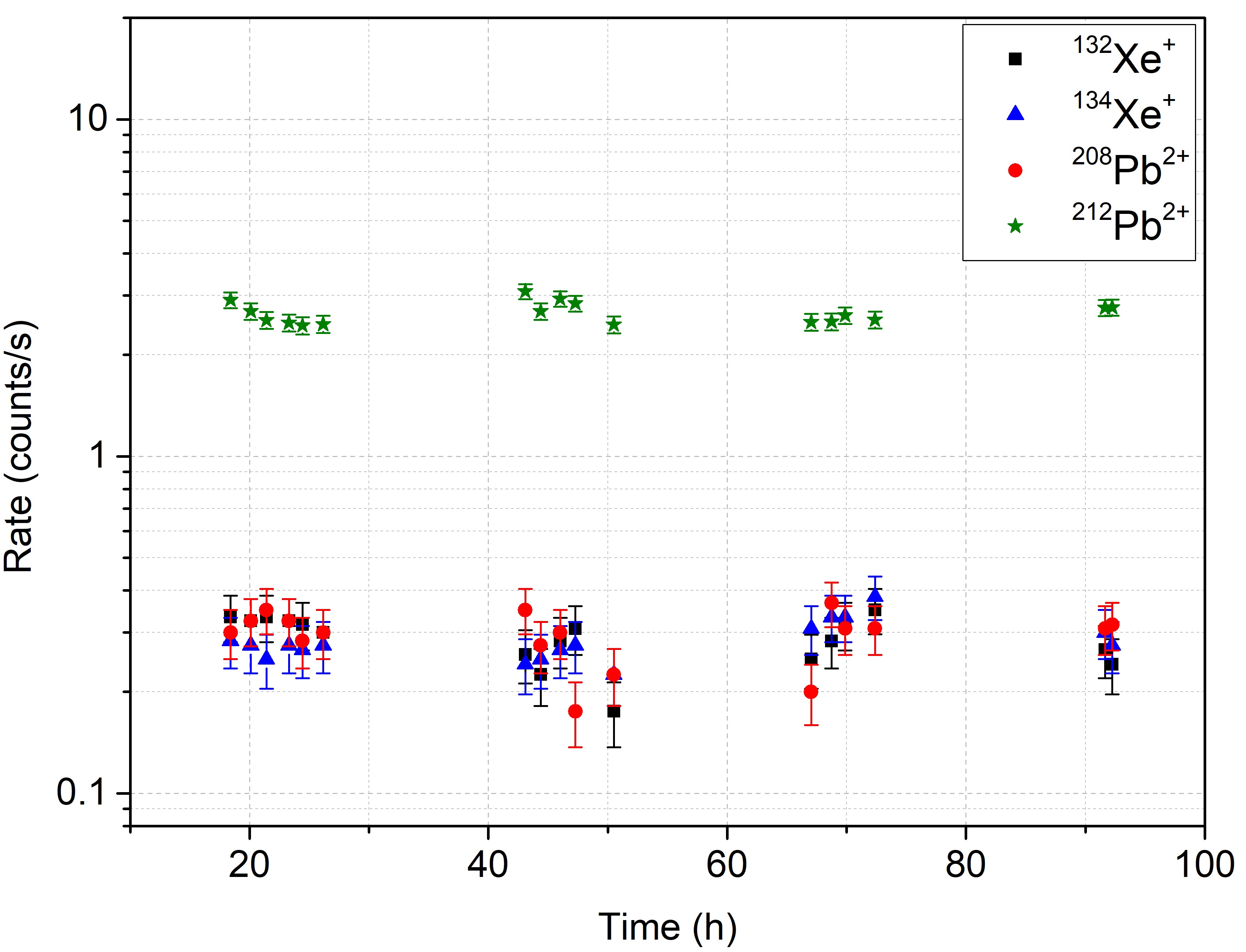}
\captionsetup{width=1.0\linewidth, justification=justified}
\caption[]{The Xe$^+$ and Pb$^{2+}$ ions measured with the MR-TOF-MS over four days of continuous operation of the CSC at 165 mbar. The cold trap was used. The Pb$^{2+}$ ions from the Th source installed inside the CSC shows the stable performance of setup while Xe contamination stabilized at a low level. 
}
\label{xe-cps}
\end{figure}

A measurement under exactly the same conditions (i.e., same gas bundle), but with the cold trap in the gas line was done over four days of continuous operation of the CSC. The CSC was filled at 165 mbar and the Xe contamination is measured after the system reached a stable condition. Figure \ref{xe-cps} shows a low and a stable contamination level for the Xe isotopes and a stable rate of Pb ions from the $^{228}$Th source. 
The resulting contamination level for $^{132}$Xe isotope after four days of operation remained stable. Without the cold trap the bundle to bundle fluctuations and the ``history'' of the CSC could add up and result in variations of the detected Xe level by two orders of magnitude. Thus, the cold trap ensures long-term low and stable detected Xe rate.  

\section{Gas purification measurements}
The CSC can be filled via four gas lines with different gas purification systems, see Figure \ref{gas-handeling}. The CSC was filled with the helium 6.0 gas: (a) directly without any gas purifier used, (b) via the passive getter (SAES Microtorr gas purifier), (c) via the passive getter and the cold trap, and (d) via the active getter (SAES Monotorr gas purifier) and the cold trap. The gas cleanliness for these different conditions was investigated with the RGA. To ensure a safe operation of the RGA, the CSC was operated at low density (temperature of 85 K and a helium pressure of 18 mbar), resulting in a pressure in the RFQ section below $5 \times 10^{-5}$ mbar (see table \ref{P} for nominal pressures in CSC and RFQ section for different modes of operation). The gas cleanliness of the CSC was monitored by using the techniques discussed in section \ref{diognostic}.

\begin{table}[]
    \centering
    \footnotesize
    \begin{tabular}{c|cc}
        Operation mode & CSC (mbar) & RFQ (mbar) \\
        \midrule
        RGA scans & 18  & 5 $\times 10^{-5}$  \\ 
        Ext. RFQ mass scans & 30  & 3 $\times 10^{-4}$  \\
        Normal CSC operations & 30-165  & (0.03-2) $\times 10^{-2}$  \\
    \end{tabular}
    \caption{Nominal pressures in CSC and RFQ section for different operation modes.}
    \label{P}
\end{table}

\begin{figure}[h!]
\centering
\stackinset{l}{2in}{b}{1.4in}{(a)}
{\hspace{0 cm}\includegraphics[width=0.78\linewidth]{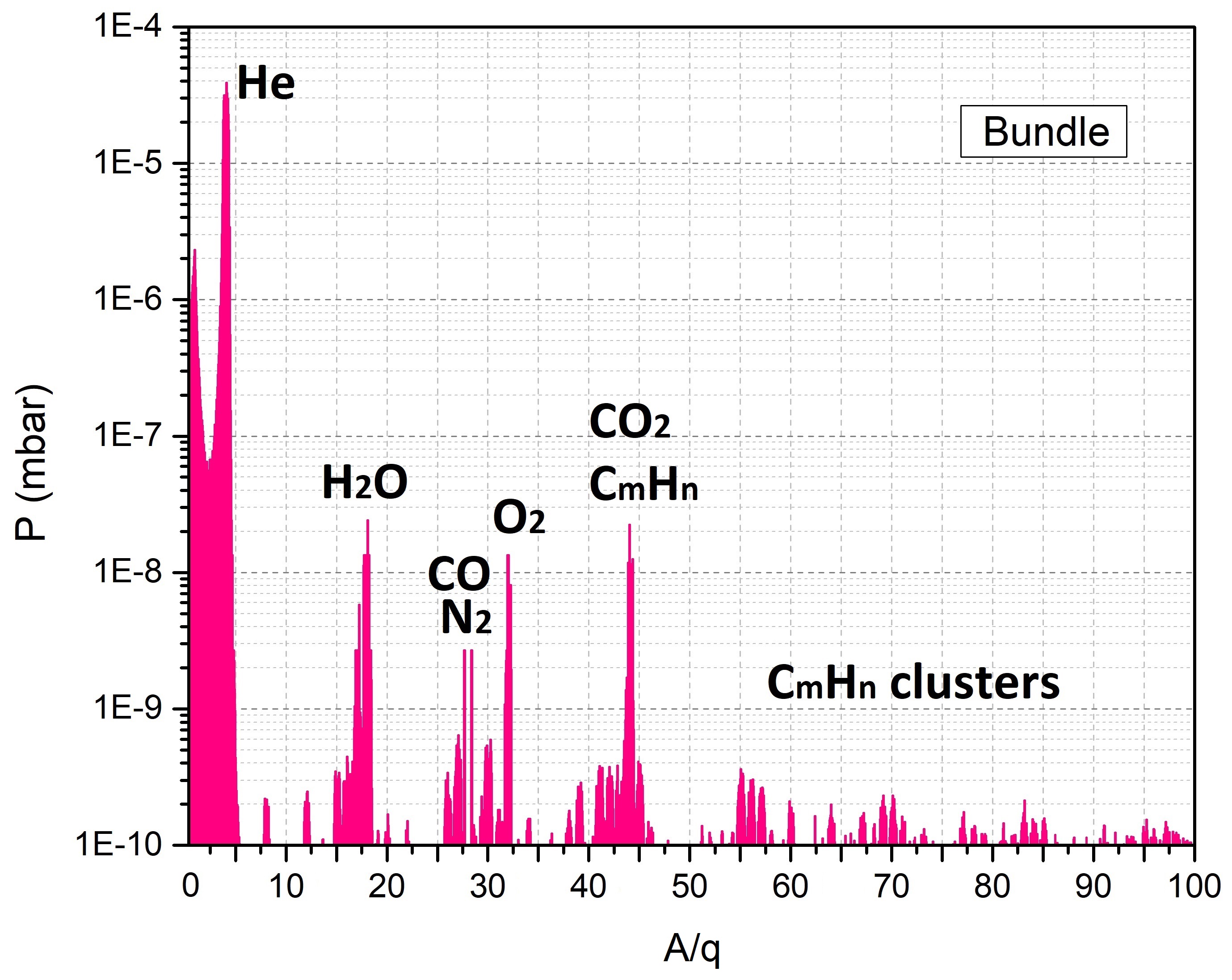}}
\centering

\stackinset{l}{2in}{b}{1.4in}{(b)}
 {\hspace{0 cm}\includegraphics[width=0.78\linewidth]{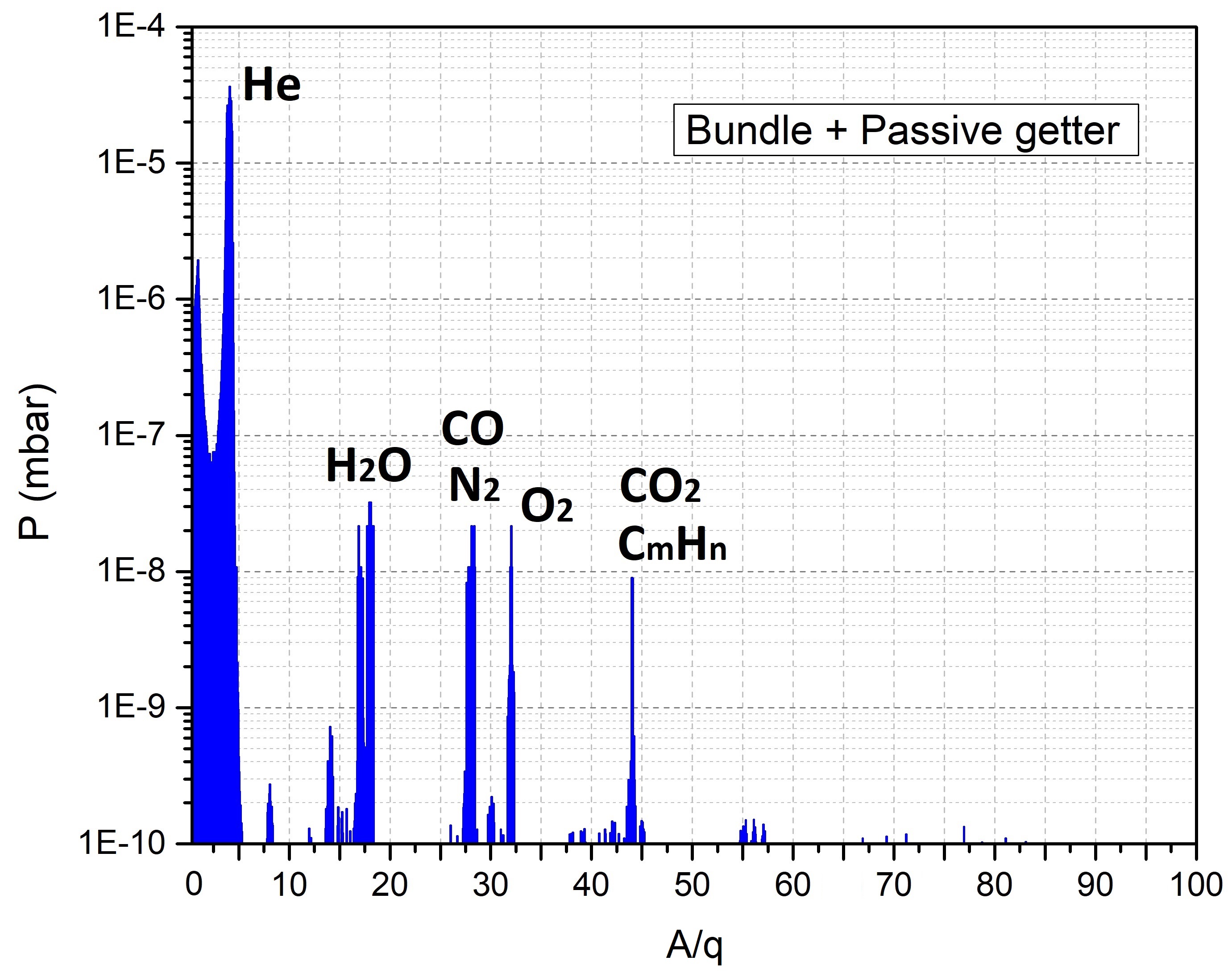}}
\centering

\stackinset{l}{2in}{b}{1.4in}{(c)}
{\hspace{0 cm}\includegraphics[width=0.78\linewidth]{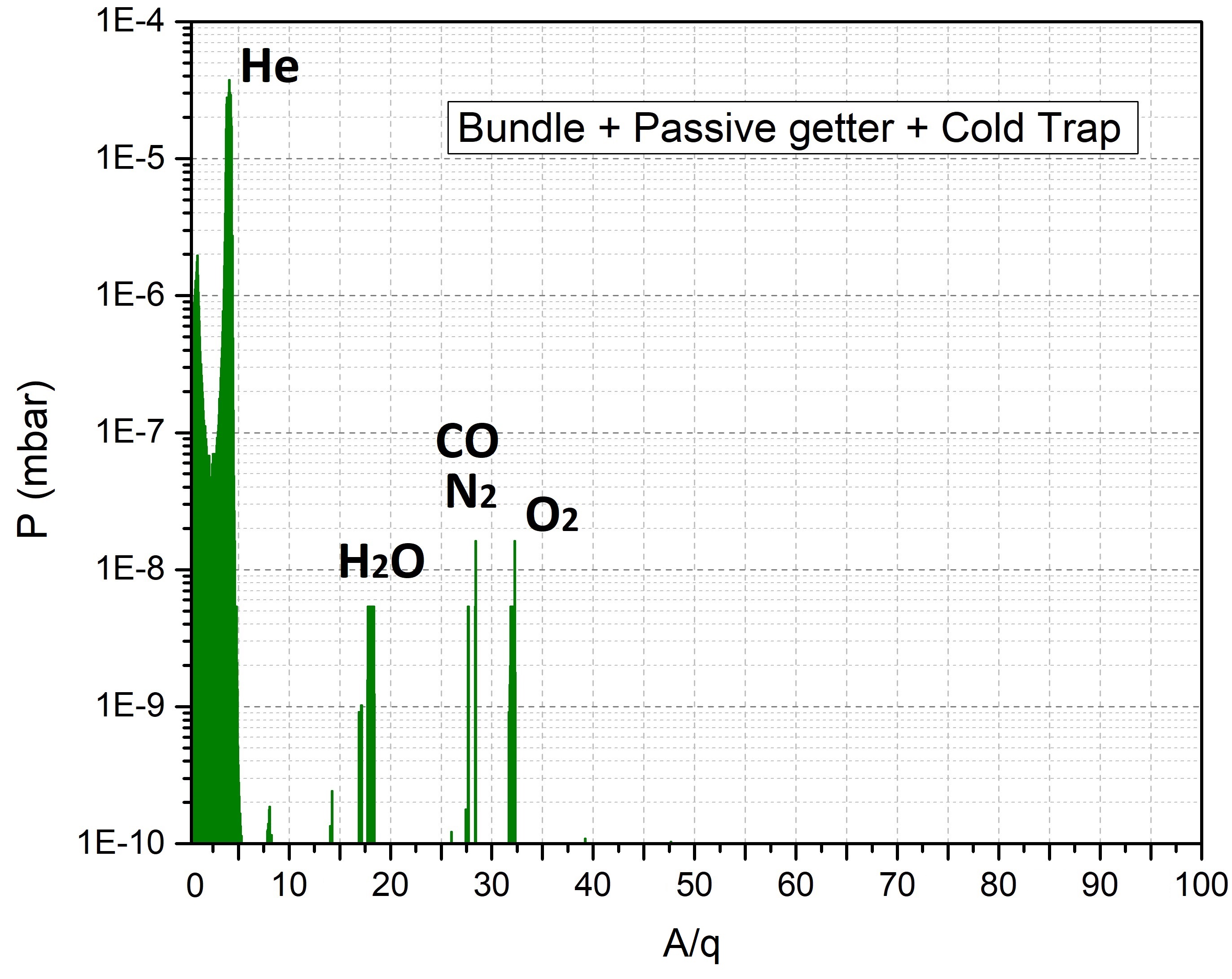}}
\centering

\stackinset{l}{2in}{b}{1.4in}{(d)}
 {\hspace{0 cm}\includegraphics[width=0.78\linewidth]{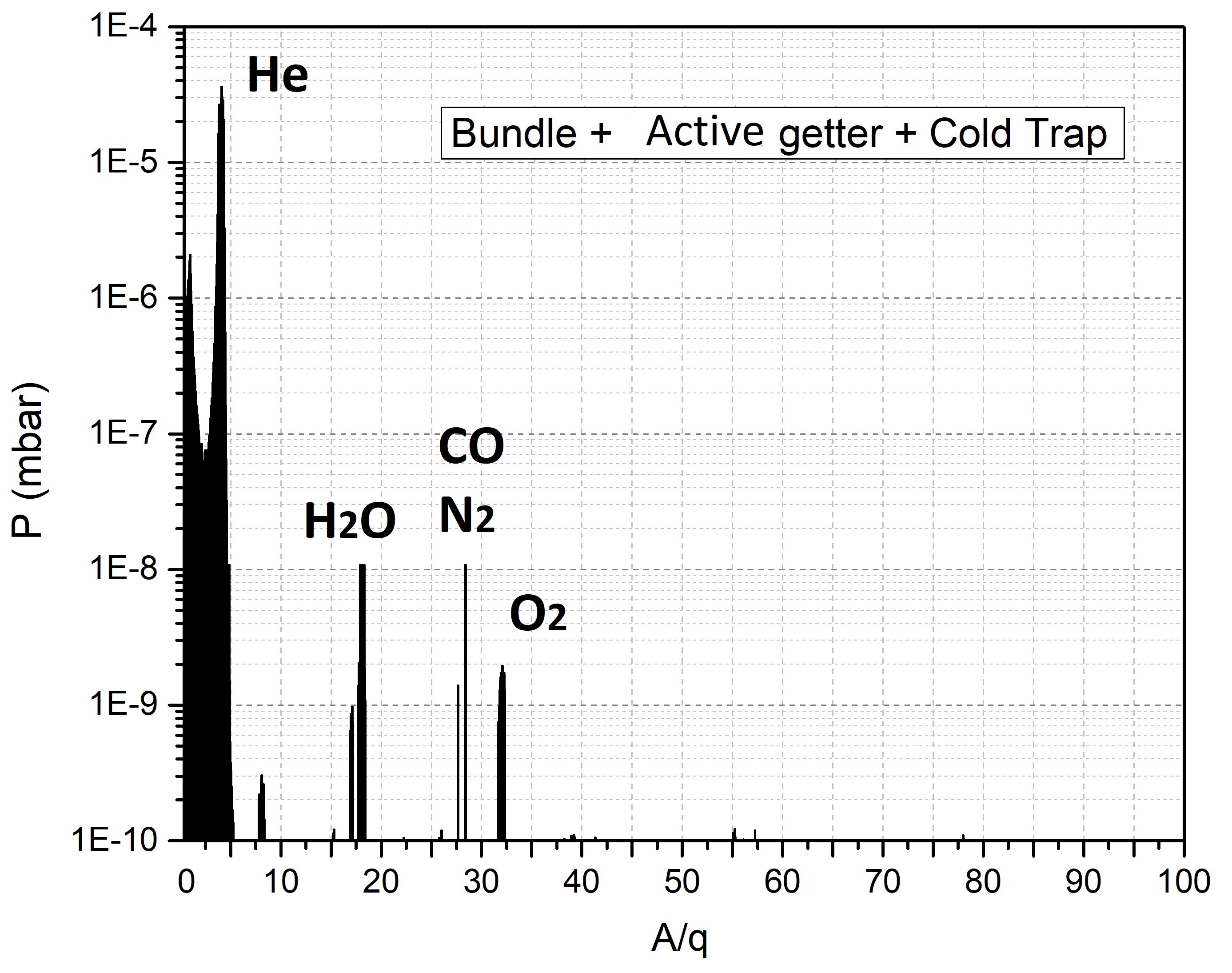}}
\captionsetup{width=1.0\linewidth, justification=justified}
\caption[]{The RGA scans for filling the CSC from four different gas purification lines. The CSC is filled with a He 6.0 gas: (a) directly, (b) via passive getter, (c) via passive getter and cold trap, and (d) via active getter and cold trap.}
\label{RGA}
\end{figure}

Figure \ref{RGA} shows the RGA scans for the four gas-filling conditions mentioned above. The dominant peaks presented in the spectra are labeled by using the molecular composition identification software of the RGA. In case (a) the dominant impurities are H$_2$O, CO, N$_2$, CO$_2$ and hydrocarbons C$_m$H$_n$. The passive getter (b) could effectively remove the heavy hydrocarbons (C$_m$H$_n$) and molecules for mass-to-charge values of $m/q=40-100$ u/e. However, this did not suppress CO$_2$ although according to the specifications of the getter \cite{SAES}, it should have been suppressed. Adding the cold trap plus passive getter (c) and plus active getter (d) significantly improved the gas cleanliness. The presence of H$_2$O in the cryogenic gas is not expected. With the measurement of the H$_2$O content going into the CSC, we conclude that the H$_2$O seen with the RGA must be from the base pressure of the RFQ beamline. Table \ref{ratio} shows the ratio between the dominant peaks in spectra compared to the He pressure.

\begin{table*}[]
    \centering
    \begin{tabular}{c|cccc}
        Configuration & m/q=18 & m/q=28 & m/q=32 & m/q=44 \\
        \midrule
       Bundle & 4.3e-4 & 7.4e-6 & 1.5e-4 & 2.2e-4\\
       Bundle + Passive getter & 8.8e-4 & 5.3e-4 & 1.7e-4 & 9.5e-5\\
       Bundle + Passive getter + Cold Trap & 1.7e-4 & 9.8e-5 & 1.6e-4 & 0\\
       Bundle + Active getter + Cold Trap & 1.5e-4 & 7e-5 & 5e-5 & 0\\
    \end{tabular}
    \caption{The ratio between the dominant peaks in Figure \ref{RGA} compared to the He pressure.}
    \label{ratio}
\end{table*}

\begin{figure}[h]
\begin{subfigure}{1\linewidth}
\centering
\phantomcaption
{\vspace{-0.5 cm}\includegraphics[width=1\linewidth]{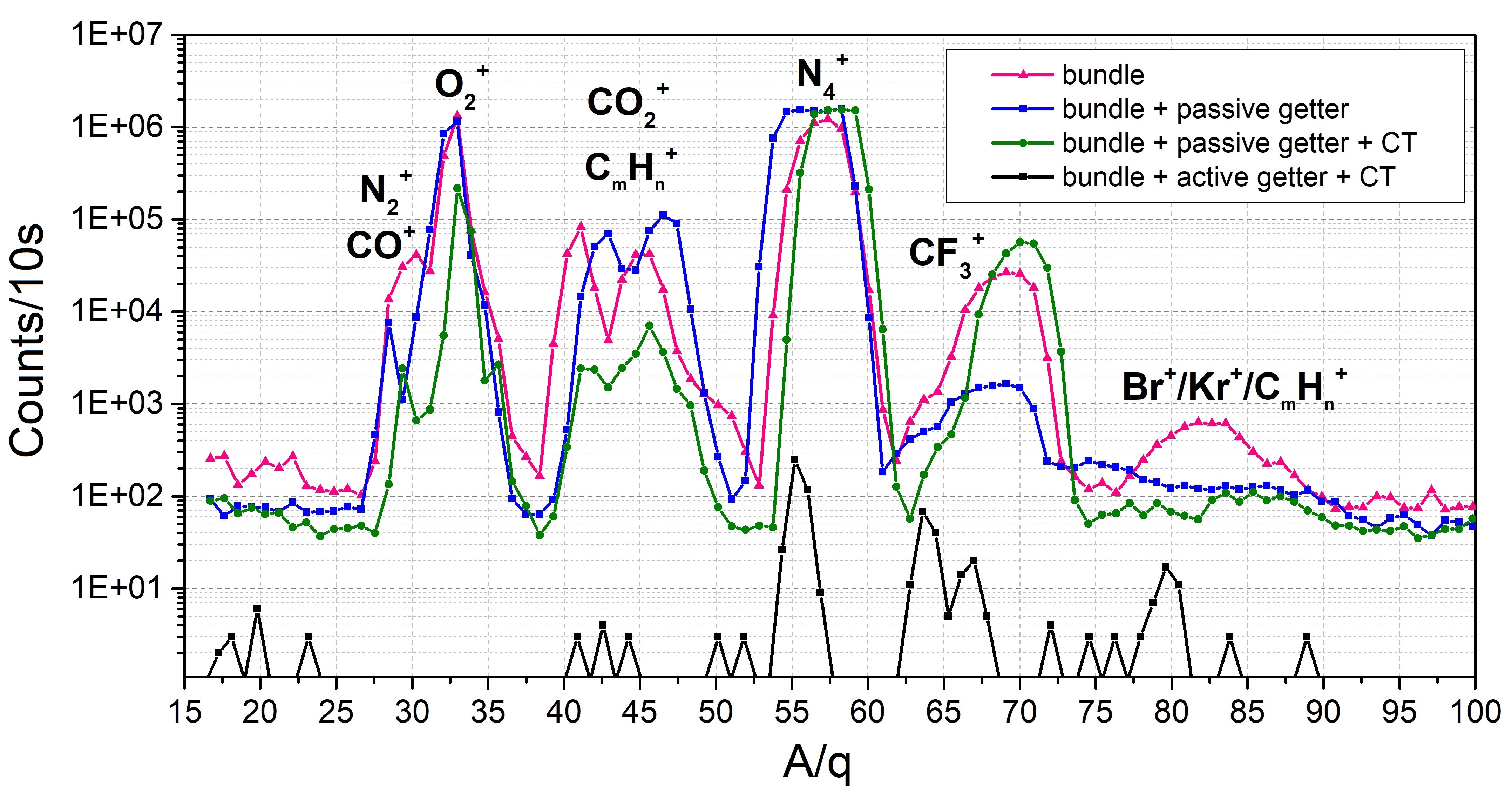}}
\end{subfigure}%
\captionsetup{width=1.0\linewidth, justification=justified}
\caption[]{The extraction-RFQ scans corresponding to the gas conditions a, b, c and d presented in Figure \ref{RGA} with the CSC filled up to 30 mbar. The impurities in the cryogenic gas are ionized by the radioactive sources installed inside the CSC (thorium alpha source and californium fission source). CT stands for the cold trap.
}
\label{chann}
\end{figure}

While the RGA scans revealed the neutral impurities in the buffer gas of the CSC, the ionized impurities formed in cryogenic gas and extracted from the CSC were investigated by using the extraction-RFQ as a mass filter for each of the gas conditions discussed above. For this, the CSC pressure was increased to 30 mbar to have a stable operation of the ion transport along the RF carpet. The CSC was pumped down to a base pressure of $5 \times 10^{-6}$ mbar after every step of the measurements before switching to a different gas purification condition. The corresponding extraction-RFQ scans are shown in Figure \ref{chann}. The mass scans are performed by scanning the DC and RF voltages over a fixed ratio (a/q=0.16). The mass spectrum covers the values of mass-to-charge from $m/q\approx$ 28 to 100 u/e. Lower mass-to-charge values ($m/q < 28$ u/e) were not effectively extracted due to the voltages set on the RF carpet and the extraction RFQ. A maximum rate of $10^6$ counts/10s can be handled by the data-acquisition system connected to the channeltron detector. The heavier impurities like N$_4^+$, CF$_3^+$, Br$^+$ and Kr$^+$ have been also identified by the MR-TOF-MS at high mass resolving powers and are used as a mass calibrants for the extraction RFQ mass scan. The count rate for N$_4^+$ in the MR-TOF-MS is many orders of magnitude lower as this molecule easily dissociates in the beamline due to its very low molecular binding energy. The ionized nitrogen cluster (N$_4^+$) is the dominant peak (not seen in the RGA scans in the neutral gas state). The N$_4^+$ is formed in ion-clustering reactions described by the two different models presented in Refs. \cite{N4,N4-second-model}.  

The passive getter (Microtorr gas purifier) used in the gas line is not capable of removing nitrogen and methane in the gas. Installing an active (heated) getter (Monotorr gas purifier) in the gas line can effectively suppress the nitrogen and the other impurities in the gas. The CF$_3^+$ contamination is expected from the Teflon sheets used inside the CSC as the high-voltage insulator. However, the fluctuation of the CF$_3^+$ peak in different scans is not understood.

The extraction-RFQ scan with the active getter in the helium gas line shows the improvement in the gas cleanliness and an effective suppression of the N$_4^+$ of almost 4 orders of magnitude. Determining the concentration level of these light contamination (N$_2$, O$_2$ and CO$_2$) similar to the heavy noble gases is not possible because the RF carpet extraction efficiency is not well known for these light masses.





\subsection{H$_2$O and O$_2$ gas monitors}

\begin{figure}[h!] 
\centering
\includegraphics[width=1\linewidth]{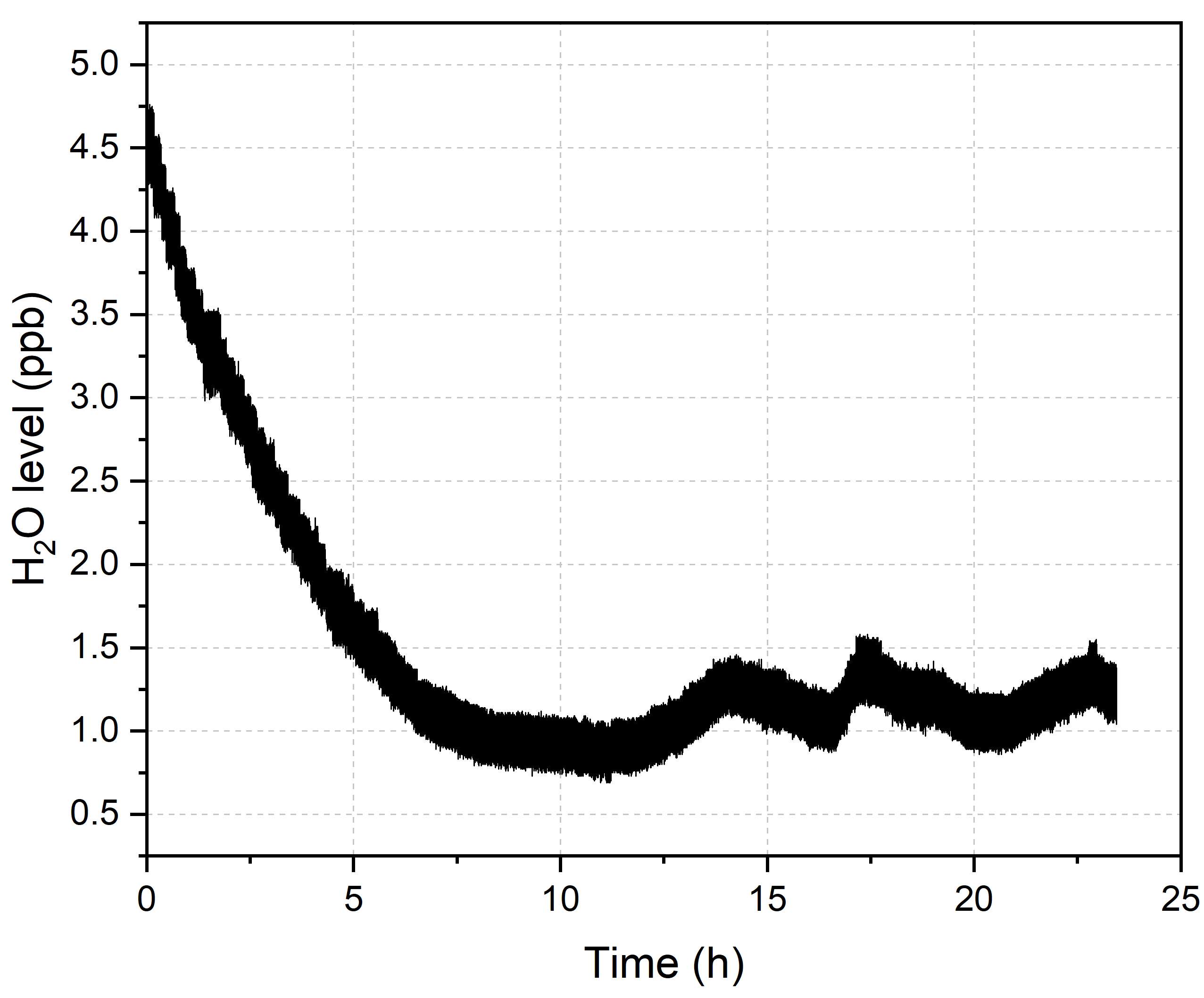}
\captionsetup{width=1.0\linewidth, justification=justified}
\caption[]{The H$_2$O level in the CSC with a helium gas pressure of up to 75 mbar filled via the cold trap and the passive getter.}
\label{H2O}
\end{figure}

Two commercial gas monitors are used for monitoring H$_2$O and O$_2$ content in the helium gas. The H$_2$O sensor (PURA OEM -120/-40 DIGS, Michell) is directly placed in the gas line and located at the closest point before gas inlet of the CSC (see Figure \ref{gas-handeling}). The H$_2$O level is measured over 23 hours of the CSC continuously filling at 75 mbar via the cold trap and the passive getter gas line. The H$_2$O level dropped from around 4.5 ppb down to 1 ppb over 7 hours (see Figure \ref{H2O}). 

The O$_2$ sensor (P12-UHP-B 50 PPB, Analytical Industries Inc.) is used for monitoring the commissioning of a new gas line. In the future, the gas bundle will be placed outside of the experimental area to transport the gas over about 100 m gas line to the FRS-IC location. The long gas line is purged with N$_2$ gas and the Oxygen level is monitored while continuously flushing the gas line with a flow rate of 1.6 standard cubic feet per hour N$_2$ gas. An O$_2$ level of 160 ppb is achieved by flushing the gas line over 16 hours without using any gas purifier. The O$_2$ sensor will be fully integrated into the gas handling system in future.


\section{Conclusions}
The gas handling system of the FRS-IC setup was upgraded with a larger capacity helium bundle, various gas distribution lines and the new gas purification system, including an active gas purifier (Monotorr). To investigate and validate gas cleanliness, new techniques for monitoring diverse types of gas contamination have been developed and applied. The new techniques make it possible to monitor the impurities both in the form of neutral or ionized particles extracted from the CSC. More than four orders of magnitude suppression of nitrogen contamination is observed by installation of the Monotorr (active) in the gas line. It is expected that this will reduce the nitrogen contamination level from 0.5 ppm (specified for a standard helium 6.0 gas bundle) to sub ppb level. Noble gases pose a special challenge as they are not removed by getters. To overcome this problem, the cold trap was used successfully for the suppression of noble gases like Xe and Kr via cryogenic adsorption. Suppression factors of 7.3 and 2.1 were achieved for Xe and Kr atoms, respectively. The measurements were used to estimate the number of expected Xe$^+$ isotopes produced in the case of measurements with beams from the FRS. The cold trap was tested for long-term operation of the CSC. It was shown that for a long operation time (4 days), the Xe level can be reduced by about two orders of magnitude. A new set of H$_2$O and O$_2$ gas sensors were added to the system. The H$_2$O sensor can detect a 1 ppb level of water vapor in the helium gas line just before entering the CSC.

The improvements now allow stable long-term conditions with increased sensitivity and accuracy in the mass measurements and more flexible operation of the cryogenic gas-filled stopping cell of the FRS Ion Catcher.
Furthermore, a new gas line was prepared to introduce trace gases into the CSC in the future to manipulate the charge-state of ions. Trace gases with low ionization potentials (IP) and high vapor pressures, like NO (IP of 9.25 eV) and CH$_4$ (IP of 12.70 eV) are planned to be used for charge-state manipulation of the extracted ions from the CSC.

\section*{Acknowledgments}
The authors acknowledge the scientific discussions with Prof. David Morrissey. This work was supported by the German Federal Ministry for Education and Research (BMBF) under contracts no.\ 05P19RGFN1 and 05P21RGFN1, by the German Research Foundation (DFG) under contract no.\ SCHE 1969/2-1, by the Hessian Ministry for Science and Art (HMWK) through the LOEWE Center HICforFAIR, by HGS-HIRe, and by Justus-Liebig-Universit{\"a}t Gie{\ss}en and GSI under the JLU-GSI strategic Helmholtzpartnership agreement. This research was supported in part by the ExtreMe Matter Institute EMMI at the GSI Helmholtzzentrum f{\"u}r Schwerionenforschung, Darmstadt, Germany, and was partly financed by the Romanian Ministry of Research, Innovation and Digitalization under the contract PN 23 21 01 06. 



 \bibliographystyle{elsarticle-num} 
 \bibliography{references}
\biboptions{numbers,sort&compress}




\end{document}